\documentclass[aps,prd,superscriptaddress,showpacs,preprint]{revtex4}
\usepackage{graphicx, bm}
\begin{document}
\draft
\title{Limits on the Electromagnetic and Weak Dipole Moments of the
       Tau-Lepton in $E_6$ Superstring Models}

\author{ A. Guti\'errez-Rodr\'{\i}guez}
\affiliation{\small Facultad de F\'{\i}sica, Universidad Aut\'onoma de Zacatecas\\
         Apartado Postal C-580, 98060 Zacatecas, Zacatecas M\'exico.\\
         Cuerpo Acad\'emico de Part\'{\i}culas Campos y Astrof\'{\i}sica.}
\author{M. A. Hern\'andez-Ru\'{\i}z}
\affiliation{\small Facultad de Ciencias Qu\'{\i}micas, Universidad Aut\'onoma de Zacatecas\\
        Apartado Postal 585, 98060 Zacatecas, Zacatecas M\'exico.\\}
\author{M. A. P\'erez}
\affiliation{\small Departamento de F\'{\i}sica, CINVESTAV.\\
             Apartado Postal 14-740, 07000, M\'exico D.F., M\'exico.}

\date{\today}

\begin{abstract}
We obtain limits on the electromagnetic and weak dipole moments of
the tau-lepton in the framework of a Left-Right symmetric model
(LRSM) and a class of $E_6$ inspired models with an additional
neutral vector boson $Z_\theta$. Using as an input the data
obtained by the L3 and OPAL Collaborations for the reaction
$e^+e^-\to \tau^+\tau^-\gamma$, we get a stringent limit on the
LRSM mixing angle $\phi$, $-1.66\times 10^{-3}< \phi<1.22\times
10^{-3}$, which in turn induces bounds on the tau weak dipole
moments which are consistent with the bounds obtained recently by
the DELPHI and ALEPH Collaborations from the reaction $e^+e^-\to
\tau^+\tau^-$. We also get similar bounds for the weak dipole
moments of the tau lepton in the framework of $E_6$ superstring
models.
\end{abstract}

\pacs{13.40.Em, 14.60.Fg, 12.15.Mm, 12.60.-i\\
Keywords: Electric and magnetic moments, taus, neutral currents, models beyond the standard model.\\
\vspace*{2cm}\noindent  E-mail: $^{1}$alexgu@planck.reduaz.mx,
$^{2}$mahernan@uaz.edu.mx, $^{3}$mperez@fis.cinvestav.mx.}

\vspace{5mm}

\maketitle

\section{Introduction}

The production of tau-lepton pairs in high energy  $e^+e^-$
collisions has been used to set bounds on its electromagnetic and
weak dipole moments \cite{Lohmann,L3D,DELPHI,ALEPH}. In the
Standard Model (SM) \cite{S.L.Glashow,S.Weinberg,A.Salam}, the
$\tau$ anomalous magnetic moment (MM) $a_\tau=(g_\tau-2)/2$ is
predicted to be $(a_\tau)_{SM}=0.0011773(3)$ \cite{Samuel,Hamzeh}
and the respective electric dipole moment (EDM) $d_\tau$ is
generated by the GIM mechanism only at very high order in the
coupling constant \cite{Barr}. Similarly, the weak MM and EDM are
induced in the SM at the loop level giving
$a^W_\tau=-(2.10+0.61i)\times 10^{-6}$ \cite{Bernabeu,Bernabeu1}
and $d^W_\tau \leq 8\times 10^{-34}e$cm \cite{Bernreuther,Booth}.
Since the current  bounds on these dipole moments
\cite{Lohmann,L3D,DELPHI,ALEPH} are well above the SM predictions,
it has been pointed out that these quantities are excellent
candidates to look for physics beyond the SM \cite{Bernabeu,
Bernabeu1,Bernreuther,Booth,Gonzalez-Garcia,Poulose,Huang,Escribano,
Grifols,Taylor,G.Gonzalez,G.Gonzalez1,G.Gonzalez2}. The couplings
of the photon and $Z$ gauge boson to charged leptons may be
parametrized in the following form

\begin{equation}
\Gamma^{\alpha}_V=eF_{1}(q^{2})\gamma^{\alpha}+\frac{ie}{2m_l}F_{2}(q^{2})\sigma^{\alpha
\mu}q_{\mu}+ eF_{3}(q^{2})\gamma_5\sigma^{\alpha \mu}q_{\mu},
\end{equation}

\noindent where $V=\gamma, Z$, $m_l$ is the lepton mass and
$q=p'-p$ is the momentum transfer. The $q^2$-dependent
form-factors $F_i(q^2)$ have familiar interpretations for $q^2=0$:
$F_1(0)\equiv Q_l$ is the electric charge; $F_2(0)\equiv a_l$; and
$F_3\equiv d_l /Q_l$. The weak dipole moments are defined in a
similar way: $F^Z_2(q^2=m^2_Z)=a^W_\tau$ and
$F^Z_3(q^2=m^2_Z)=d^W_\tau/e$. The measurement of $a^W_\tau$ and
$d^W_\tau$ has been done in the $Z_1 \to \tau^+ \tau^-$ decay mode
at LEP. The latest bounds obtained for the electromagnetic and
weak dipole moments from the DELPHI and ALEPH collaborations at
the 95$\%$ C.L. are: $-0.052 < a_\tau < 0.013$, $-0.22 < d_\tau
(10^{-16}\hspace{1mm} e\mbox{cm}) < 0.45$ and $a^W_\tau <
1.1\times 10^{-3}$, $d^W_\tau < 0.50\times 10^{-17}e\mbox{cm}$
\cite{DELPHI,ALEPH}.

The first limits on the MM and EDM of the $\tau$ lepton were
obtained by Grifols and M\'endez using L3 data \cite{Grifols}:
$a_\gamma\leq 0.11$ and $d_\gamma\leq 6\times 10^{-16}e$cm.
Escribano and Mass\'o \cite{Escribano} later on used electroweak
precision measurements to get $d_\tau \leq 1.1\times 10^{-17}e$cm
and $-0.004\leq a_\tau \leq 0.006$ at the $2\sigma$ confidence
level. There is an extensive theoretical work done in models
beyond the SM that contribute to EDM of charged leptons. In Ref.
\cite{Iltan} the EDM of charged leptons are studied assuming that
they have Gaussian profiles in extra dimensions. In \cite{Dutta}
the lepton EDM has been analyzed in the framework of the seesaw
model. The electric dipole moments of the leptons in the version
III of the 2HDM are considered in \cite{Iltan1}. The work
\cite{Iltan2} was related to the lepton EDM in the framework of
the SM with the inclusion of non-commutative geometry.
Furthermore, the effects of non-universal extra dimensions on the
EDM of fermions in the two Higgs doublet model have been estimated
in Ref. \cite{Iltan3}.

The existence of a heavy neutral ($Z'$) vector boson is a feature
of many extensions of the standard model. In particular, one (or
more) additional $U(1)'$ gauge factor provides one of the simplest
extensions of the SM. Additional $Z'$ gauge bosons appear in Grand
Unified Theories (GUT's)
\cite{Robinett,Robinett1,Robinett2,Langacker2}, Superstring
Theories
\cite{Green,Green1,Gross,Witten,Witten1,Candelas,Dine,Ellis,Breit,Candelas1,Cecotti},
Left-Right Symmetric Models (LRSM)
\cite{Mohapatra,G.Senjanovic,Shrock,G.Senjanovic1} and in other
models such as models of composite gauge bosons
\cite{Schrempp,Baur,Kuroda}. The largest set of extended gauge
theories are those which are based on GUT's. Popular examples are
the groups $SO(10)$ and $E_6$. Generically, additional $Z$-bosons
originating from $E_6$ grand unified theories are conveniently
labeled in terms of the chain: $E_6 \to SO(10)\times U(1)_\psi \to
SU(5)\times U(1)_\chi \times U(1)_\psi \to SM \times
U(1)_{\theta_{E_6}}$ where $U(1)_{\theta_{E_6}}$ remains unbroken
at low energies. Detailed discussions on GUTS can be found in the
literature
\cite{Robinett,Robinett1,Robinett2,Langacker2,Green,Green1,Gross,Witten,Witten1,Candelas,Dine,Ellis,Breit,Candelas1,Cecotti}.

On the $Z_1$ peak, where a large number of $Z_1$ events are
collected at $e^{+}e^{-}$ colliders, one may hope to constrain or
eventually measure the electromagnetic and weak dipole moments of
the $\tau$ by selecting $\tau^{+}\tau^{-}$ events accompanied by a
hard photon. The Feynman diagrams which give the most important
contribution to the cross section from $e^{+}e^{-}\rightarrow
\tau^+ \tau^- \gamma$ are shown in Fig. 1. The total cross section
of $e^{+}e^{-}\rightarrow \tau^+ \tau^- \gamma$ will be evaluated
at the $Z_1$-pole in the framework of a left-right symmetric model
and a class of $E_6$ inspired models. The numerical computation
for the anomalous magnetic and the electric dipole moments of the
tau is done using the data collected by the L3 and OPAL
Collaborations at LEP \cite{L3,OPAL}. We are interested in
studying the effects induced by the effective couplings associated
to the weak and electromagnetic moments of the tau lepton given in
Eq. (1). For this purpose we will take the respective anomalous
vertices $\tau\tau\gamma$ and $\tau\tau Z$, one at the time, in
diagrams (1) and (2) of Fig. 1. The numerical computation for the
respective transition amplitudes will be done using the data
collected by these collaborations.

Our aim in this paper is to analyze the reaction
$e^{+}e^{-}\rightarrow \tau^+ \tau^- \gamma$ in the $Z_1$ boson
resonance. The analysis is carried out in the context of a
left-right symmetric model \cite{A.Gutierrez,A.Gutierrez3} and a
class of $E_6$ inspired models with an additional neutral vector
boson $Z_\theta$ \cite{Aytekin,Aydin} and we attribute a weak and
electromagnetic dipole moments to the tau lepton. Processes
measured in the resonance serve to set limits on the tau
electromagnetic and weak dipole moments. First, using as an input
the results obtained by the L3 and OPAL Collaborations
\cite{L3,OPAL} for the tau MM and EDP in the process $e^+e^-\to
\tau^+\tau^-\gamma$, we will set a limit on the LRSM mixing angle
$\phi$ which is similar to that obtained recently from the LEP
data on the number of light neutrino species \cite{A.Gutierrez2}
but it is stronger by about one order of magnitude than the
previous limit for this angle obtained from old LEP data
\cite{M.Maya}. We then use this limit on $\phi$ to get bounds on
the weak dipole moments of the tau from the same L3/OPAL data. We
have found that these limits are consistent with the new bounds
obtained by the DELPHI and ALEPH Collaborations from the process
$e^+e^-\to \tau^+\tau^-$ \cite{L3E,DELPHI,ALEPH}. We will get also
similar bounds for these dipole moments using the known limits for
the $E_6$ mixing angle $\phi_{(SM-E_6)}$ between $Z_1$ and
$Z_\theta$ \cite{London,Capstick}.

This paper is organized as follows: In Sect. II we describe the
neutral current couplings in $E_6$. In Sect. III we present the
calculation of the cross section for the process
$e^{+}e^{-}\rightarrow \tau^+ \tau^- \gamma$. In Sect. IV we
present our results for the numerical computations and, finally,
we present our conclusions in Sect. V.

\section{Neutral Current Couplings in $E_6$}

In this section we describe the neutral current couplings involved
in the class of $E_6$ inspired models we are interested in. Let us
consider the following breakdown pattern in $E_6$:

\begin{equation}
E_6 \to SO(10)\times U(1)_\psi \to SU(5)\times U(1)_\chi \times
U(1)_\psi \to SM \times U(1)_{\theta_{E_6}},
\end{equation}

\noindent where the $SU(3)_C \times SU(2)_L \times U(1)_{Y_W}$
groups of the standard model are embeded in the $SU(5)$ subgroup
of $SO(10)$. The couplings of the fermions to the standard model
$Z_1$ are given, as usual, by

\begin{equation}
Q=I_{3L}-Q_{em}\sin^2\theta_W,
\end{equation}

\noindent while the couplings to the $Z_\theta$ are given by
linear combinations of the $U(1)_\chi$ and $U(1)_\psi$ charges
\cite{London,Capstick}:

\begin{eqnarray}
Q'&=&Q_\chi \cos\theta_{E_6}+Q_\psi \sin\theta_{E_6}, \nonumber\\
Q''&=&-Q_\chi \sin\theta_{E_6}+Q_\psi \cos\theta_{E_6},
\end{eqnarray}

\noindent where the operators $Q_\psi$ and $Q_\chi$ are orthogonal
to those of $Q_{em}$ and that of the standard model $Z_1$ and
$\theta_{E_6}$ is the $Q_\chi-Q_\psi$ mixing angle in $E_6$.

With the extra $Z_\theta$ neutral vector boson the neutral current
Lagrangian is \cite{Leike,Alam}

\begin{equation}
{-\cal L}_{NC}=eA^\mu J_{em\mu}+g_1Z^\mu_1
J_{Z_{1}\mu}+g_2Z^\mu_\theta J_{Z_{\theta}\mu},
\end{equation}

\noindent where $J_{em \mu}$, $J_{Z_{1}\mu}$ and
$J_{Z_{\theta}\mu}$ are the electromagnetic current, the $Z_1$
current of the standard model and the $J_{Z_\theta\mu}$ current of
the new boson, respectively and are given by

\begin{eqnarray}
J_{Z_1 \mu}&=&\sum_f\bar f\gamma_\mu(C^1_V+C^1_A\gamma_5)f, \nonumber\\
J_{Z_\theta \mu}&=&\sum_f\bar f\gamma_\mu(C'_V+C'_A\gamma_5)f,
\end{eqnarray}

\noindent where $f$ represents fermions, while

\begin{eqnarray}
g_1&=&(g^2+g'^2)^{1/2}= \frac{e}{2\sin\theta_W\cos\theta_W}, \nonumber\\
g_2&=&g_{\theta},\\
C^{1e}_V&=&-\frac{1}{2}+2\sin^2 {\theta_W}, \hspace{1cm} C^{1e}_A=-\frac{1}{2},\nonumber\\
C^{1\tau}_V&=&-\frac{1}{2}+2\sin^2 {\theta_W},\hspace*{1cm} C^{1\tau}_A=-\frac{1}{2},\\
C^{'e}_V&=&z^{1/2}(\frac{\cos\theta_{E_6}}{\sqrt{6}}+\frac{\sin\theta_{E_6}}{\sqrt{10}}),\hspace{9mm} C^{'e}_A=2z^{1/2}\frac{\sin\theta_{E_6}}{\sqrt{10}}, \nonumber\\
C^{'\tau}_V&=&z^{1/2}(\frac{\cos\theta_{E_6}}{\sqrt{6}}+\frac{\sin\theta_{E_6}}{\sqrt{10}}),\hspace*{9mm}
C^{'\tau}_A=2z^{1/2}\frac{\sin\theta_{E_6}}{\sqrt{10}},
\end{eqnarray}

\noindent with

\begin{equation}
z=(\frac{g^2_\theta}{g^2+g'^2})(\frac{M_{Z_1}}{M_{Z_\theta}})^2,
\end{equation}

\noindent a parameter that depends on the coupling constant
$g_\theta$ and $M_{Z_\theta}$.

The class of $E_6$ models we shall be interested in arise with the
following specific values for the mixing angle $\theta_{E_6}$
\cite{London}:

\begin{eqnarray}
\theta_{E_6}&=&0^o, \hspace{6mm} Z_{\theta_{E_6}}\to Z_\psi,\nonumber\\
\theta_{E_6}&=&37.8^o, \hspace{6mm} Z_{\theta_{E_6}}\to Z', \nonumber\\
\theta_{E_6}&=&90^o, \hspace{6mm} Z_{\theta_{E_6}}\to Z_\chi,\\
\theta_{E_6}&=&127.8^o, \hspace{6mm} Z_{\theta_{E_6}}\to
Z_I,\nonumber
\end{eqnarray}

\noindent where $Z_\psi$ is the extra neutral gauge boson arising
in $E_6 \to SO(10)\times U(1)_\psi$, $Z'$ corresponds to the
respective neutral gauge boson obtained if $E_6$ is broken down to
a rank-5 group, $Z_\chi$ is the neutral gauge boson involved in
$SO(10) \to SU(5)\times U(1)_\chi$, and $Z_I$ is the neutral gauge
boson associated to the breaking of $E_6$ via a non-Abelian
discrete symmetry to a rank-5 group \cite{London}.

The $Z_\theta$ can also mix with the standard model $Z_1$ so that
the physical fields, $Z_{1p}$ and $Z_{\theta p}$ are linear
combinations of the gauge fields $Z_{1n}$ and $Z_{\theta n}$ with
mixing angle $\phi_{(SM-E_6)}$. This mixing will alter the fermion
couplings to the $Z_1$ as is indicated in Eqs. (3a) and (3b) of
Ref. \cite{Capstick}. Since $\phi_{(SM-E_6)}$ comes from the
diagonalization of the $Z_1-Z_\theta$ mass matrix, it can be
expressed in terms of the standard model prediction for the $Z_1$
mass, $M_{SM}=M_W\cos^2\theta_W$ and the physical $Z_1$ and
$Z_\theta$ masses \cite{Langacker}:

\begin{equation}
\tan^2\phi_{(SM-{E_6})}=\frac{M^2_{SM}-M^2_{Z_1p}}{M^2_{Z_\theta
p}- M^2_{Z_1p}}.
\end{equation}

The bounds of the MM and EDM of the tau lepton, with the
modification of the fermion couplings to the $Z_1$ to incorporate
the mixing of $Z_1$ and $Z_\theta$ for different values of the
mixing angle $\phi_{(SM-E_6)}$ of the $E_6$ model are given in
section IV.

\section{The Total Cross Section}

We will take advantage of our previous work on the LRSM and we
will calculate the total cross section for the reaction
$e^{+}e^{-}\rightarrow \tau^+\tau^- \gamma$ using the transition
amplitudes given in Eqs. (21) and (22) of Ref. \cite{A.Gutierrez}
for the LRSM for diagrams 1 and 2 of Fig. 1. For the contribution
coming from diagrams 3 and 4 of Fig. 1, we use Eqs. (6) and (9)
given in section II for the $E_6$ model. The respective transition
amplitudes are thus given by

\begin{eqnarray}
{\cal M}_{1}&=&\frac{-g^{2}}{4\cos^{2}\theta_{W}(l^{2}-m^{2}_{\tau})}[\bar u(p_{3})\Gamma^{\alpha}(l\llap{/}+m_{\tau})\gamma^{\beta}(aC^{1\tau}_V-bC^{1\tau}_A \gamma_{5})v(p_{4})]\nonumber\\
         &&\frac{(g_{\alpha\beta}-p_{\alpha}p_{\beta}/M^{2}_{Z_1})}{[(p_{1}+p_{2})^{2}-M^{2}_{Z_1}-i\Gamma^{2}_{Z_1}]}[\bar u(p_{2})\gamma^{\alpha}(aC_V^{1e}-bC_A^{1e}\gamma_{5})v(p_{1})]\epsilon^{\lambda}_{\alpha},
\end{eqnarray}


\begin{eqnarray}
{\cal M}_{2}&=&\frac{-g^{2}}{4\cos^{2}\theta_{W}(k^{2}-m^{2}_{\tau})}[\bar u(p_{3})\gamma^{\beta}(aC^{1\tau}_V-bC^{1\tau}_A\gamma_{5})(k\llap{/}+m_{\tau})\Gamma^{\alpha}   v(p_{4})]\nonumber\\
         &&\frac{(g_{\alpha\beta}-p_{\alpha}p_{\beta}/M^{2}_{Z_1})}{[(p_{1}+p_{2})^{2}-M^{2}_{Z_1}-i\Gamma^{2}_{Z_1}]}[\bar u(p_{2})\gamma^{\alpha}(aC_V^{1e}-bC_A^{1e}\gamma_{5})v(p_{1})]\epsilon^{\lambda}_{\alpha},
\end{eqnarray}

\noindent and for $M'_1$ and $M'_2$

\begin{eqnarray}
M'_1&=&M_1(aC^1_V \to C_V', bC^1_A \to C_A', M_{Z_1} \to M_{Z_\theta}),\\
M'_2&=&M_2(aC^1_V \to C_V', bC^1_A \to C_A', M_{Z_1} \to
M_{Z_\theta}),
\end{eqnarray}

\noindent where $\Gamma^{\alpha}$ is the tau-lepton
electromagnetic vertex which is defined in the Eq. (1), while
$\epsilon^{\lambda}_{\alpha}$ is the polarization vector of the
photon. $l$ ($k$) stands for the momentum of the virtual tau
(antitau), and the coupling constants $a$ and $b$ are given in the
Eq. (16) of the Ref. \cite{A.Gutierrez}, while $C'_V$ and $C'_A$
are given above in the Eq. (9).

The MM, EDM, the mixing angle $\phi$ of the LRSM as well as the
mixing angle $\theta_{E_6}$ and the mass of the additional neutral
vector boson $M_{Z_\theta}$ of the $E_6$ model give a contribution
to the differential cross section for the process
$e^{+}e^{-}\rightarrow \tau^+\tau^-\gamma$ of the form:

\begin{eqnarray}
\frac{d\sigma}{E_{\gamma}dE_{\gamma}d\cos\theta_{\gamma}}
&=&\frac{\alpha^{2}}{48\pi}\; \{[\frac{e^2 a^{2}_\tau}{4m
^2_\tau}+d^{2}_\tau]\; [{\cal C}[\phi, x_{W}]\;
{\cal F}[\phi, s, E_{\gamma}, \cos\theta_{\gamma}]\nonumber \\
&+&{\cal C}_1[\theta_{E_6}, M_{Z_\theta}, x_{W}]\; {\cal
F}_1[M_{Z_\theta}, s, E_{\gamma}, \cos\theta_{\gamma}]
+8f[M_{Z_1}, \Gamma_{Z_1}, M_{Z_\theta}, \Gamma_{Z_\theta}]\nonumber\\
&\cdot&\{ {\cal C}_2[\phi, \theta_{E_6}, M_{Z_\theta}, x_{W}]\;
{\cal F}_2[s] + {\cal C}_3[\phi, \theta_{E_6}, M_{Z_\theta},
x_{W}]\; {\cal F}_3[s, E_{\gamma}] + {\cal C}_4[\theta_{E_6},
M_{Z_\theta}]\;
{\cal F}_4[E_{\gamma}]\nonumber\\
&+& {\cal C}_5[\phi, \theta_{E_6}, M_{Z_\theta}, x_{W}]\; {\cal
F}_5[s, E_{\gamma},\cos\theta_{\gamma}] + {\cal C}_6[\phi,
\theta_{E_6}, M_{Z_\theta}, x_{W}]\; {\cal F}_6[s,
E_{\gamma},\cos\theta_{\gamma}]\}]\nonumber\\
&+&\frac{3}{32}[\frac{ea_\tau}{2m_\tau}+d_\tau]\; f[M_{Z_1},
\Gamma_{Z_1}, M_{Z_\theta}, \Gamma_{Z_\theta}]\cdot \{ {\cal
C}_7[\phi, M_{Z_\theta}, x_{W}]\; {\cal
F}_7[s, E_\gamma, \cos\theta_{\gamma}]\nonumber\\
&-&{\cal C}_7[\phi, M_{Z_\theta}, x_{W}]\; {\cal F}_8[s,
E_{\gamma}, \cos\theta_{\gamma}]-{\cal C}_8[\phi, M_{Z_\theta},
x_{W}]\; {\cal F}_7[s, E_\gamma, \cos\theta_{\gamma}]\nonumber\\
&+&{\cal C}_8[\phi, M_{Z_\theta}, x_{W}]\; {\cal F}_8[s,
E_{\gamma}, \cos\theta_{\gamma}]\}+ \frac{1}{36} {\cal
C}_9[x_{W}]\; {\cal F}_9[s, E_{\gamma}, \cos\theta_{\gamma}] \},
\end{eqnarray}

\noindent where $E_{\gamma}$, $\cos\theta_{\gamma}$ are the energy
and the opening angle of the emmited photon.\\

The kinematics is contained in the functions

\begin{eqnarray}
{\cal F}[\phi, s, E_{\gamma}, \cos\theta_{\gamma}] &\equiv
&\frac{[[\frac{1}{2}(a^{2}+b^{2})-4a^2x_W+8a^2x^2_W]
(s-2\sqrt{s}E_{\gamma})+\frac{1}{2}b^{2}E^{2}_{\gamma}\sin^{2}\theta_{\gamma}]}
{[(s-M_{Z_1}^2)^2+M^{2}_{Z_1}\Gamma^{2}_{Z_1}]},\nonumber\\
{\cal F}_1[M_{Z_\theta}, s, E_{\gamma}, \cos\theta_{\gamma}]
&\equiv & \frac{[[(C^{'\tau}_V)^2+(C^{'\tau}_A)^2]
(s-2\sqrt{s}E_{\gamma})+(C^{'\tau}_A)^2
E^{2}_{\gamma}\sin^{2}\theta_{\gamma}]}
{[(s-M_{Z_\theta}^2)^2+M^{2}_{Z_\theta}\Gamma^{2}_{Z_\theta}]},\nonumber\\
{\cal F}_2[s]
&\equiv & 4\sqrt{s},\nonumber\\
{\cal F}_3[s, E_{\gamma}]
&\equiv & 2\sqrt{s}E_{\gamma},\nonumber\\
{\cal F}_4[E_{\gamma}]
&\equiv & \sqrt{15}E_{\gamma},\nonumber\\
{\cal F}_5[s, E_{\gamma}, \cos\theta_{\gamma}]
&\equiv & -(s+\frac{1}{2}E^{2}_{\gamma}\sin^{2}\theta_{\gamma}),\nonumber\\
{\cal F}_6[s, E_{\gamma}, \cos\theta_{\gamma}] &\equiv &
(s-E_{\gamma}^{2}+\frac{1}{2}E^{2}_{\gamma}\sin^{2}\theta_{\gamma}),\\
{\cal F}_7[s, E_{\gamma}, \cos\theta_{\gamma}] &\equiv &
(12\sqrt{s}E_{\gamma}+9E^{2}_{\gamma}\sin^{2}\theta_{\gamma}),\nonumber\\
{\cal F}_8[s, E_{\gamma}, \cos\theta_{\gamma}] &\equiv &
(9s+4\sqrt{s}E_{\gamma}\sin^{2}\theta_{\gamma}),\nonumber\\
{\cal F}_9[s, E_{\gamma}, \cos\theta_{\gamma}]&\equiv&
\frac{[4\sqrt{s}-2E_{\gamma}\sin^{2}\theta_{\gamma}-\sqrt{s}\sin^{2}\theta_{\gamma}]}
{[(s-M_{Z_1}^2)^2+M^{2}_{Z_1}\Gamma^{2}_{Z_1}]},\nonumber
\end{eqnarray}
\vspace{-1.2cm}
\begin{eqnarray*}
&&f(s, M_{Z_1}, \Gamma_{Z_1}, M_{Z_\theta}, \Gamma_{Z_\theta})\nonumber \\
&\equiv &
\frac{-2[(s-M^2_{Z_{1}})(s-M^2_{Z_{\theta}})+M_{Z_{1}}\Gamma_{Z_{1}}M_{Z_{\theta}}\Gamma_{Z_{\theta}}]}
{[(s-M^2_{Z_{1}})(s-M^2_{Z_{\theta}})+M_{Z_{1}}\Gamma_{Z_{1}}M_{Z_{\theta}}\Gamma_{Z_{\theta}}]^2
+[(s-M^2_{Z_{\theta}})M_{Z_{1}}\Gamma_{Z_{1}}-(s-M^2_{Z_{1}})M_{Z_{\theta}}\Gamma_{Z_{\theta}}]^2}.
\end{eqnarray*}

The coefficients ${\cal C}$, ${\cal C}_1$,..., ${\cal C}_9$ are
given by

\begin{eqnarray}
{\cal C}[\phi, x_{W}] &\equiv
&\frac{[\frac{1}{2}(a^{2}+b^{2})-4a^{2}x_{W}
+ 8a^{2}x^{2}_{W}]}{x^{2}_{W}(1 - x_{W})^{2}},\nonumber\\
{\cal C}_1[\theta_{E_6}, M_{Z_\theta}, x_{W}]
&\equiv &\frac{3[(C^{'e}_V)^{2}+ (C^{'e}_A)^{2}]}{x^{2}_{W}(1 - x_{W})^{2}},\nonumber\\
{\cal C}_2[\phi, \theta_{E_6}, M_{Z_\theta}, x_{W}]
&\equiv&\frac{[2ax_{W}-\frac{1}{2}(a-b)][C_V^{'e}C_A^{'e}-(C_A^{'e})^{2}]}{x^{2}_{W}(1-x_{W})^{2}},\nonumber\\
{\cal C}_3[\phi, \theta_{E_6}, M_{Z_\theta}, x_{W}]
&\equiv&\frac{[4ax_{W}-a][3(C_A^{'e})^{2}-(C_V^{'e}-C_A^{'e})^{2}]}{x^{2}_{W}(1-x_{W})^{2}}\nonumber\\
{\cal C}_4[\theta_{E_6}, M_{Z_\theta}, x_{W}]
&\equiv&\frac{[3(C_A^{'e})^{2}+(C_V^{'e}-C_A^{'e})^{2}]}{x^{2}_{W}(1-x_{W})^{2}},\\
{\cal C}_5[\phi, \theta_{E_6}, M_{Z_\theta}, x_{W}]
&\equiv&\frac{[3(C_A^{'e})^{2}(4ax_{W}-a+\frac{1}{2})+
[4ax_{W}-(a-b)][C_V^{'e}C_A^{'e}-(C_A^{'e})^{2}]]}{x^{2}_{W}(1-x_{W})^{2}},\nonumber\\
{\cal C}_6[\phi, \theta_{E_6}, M_{Z_\theta}, x_{W}]
&\equiv&\frac{[2ax_{W}-\frac{1}{2}a](C_V^{'e}-C_A^{'e})^{2}}{x^{2}_{W}(1-x_{W})^{2}},\nonumber\\
{\cal C}_7[\phi, M_{Z_\theta}, x_{W}]
&\equiv&\frac{[\frac{1}{2}(a^2+b^2)-4a^2x_{W}+8a^2x^2_W]
[\frac{1}{2}(a+b)-4ax_{W}+8ax^2_W]}{x^{2}_{W}(1-x_{W})^{2}},\nonumber\\
{\cal C}_8[\phi, \theta_{E_6}, M_{Z_\theta}, x_{W}]
&\equiv&\frac{[\frac{1}{2}(a^2+b^2)-4a^2x_{W}+8a^2x^2_W]
[(1-4x_{W})C_V^{'\tau}-C_A^{'\tau}]}{x^{2}_{W}(1-x_{W})^{2}},\nonumber\\
{\cal C}_9[x_{W}] &\equiv&
\frac{[1-4x_{W}+8x^2_W]^2}{x^{2}_{W}(1-x_{W})^{2}},\nonumber
\end{eqnarray}

\noindent with $x_{W}\equiv \sin^{2}\theta_{W}$.\\

In the above expressions, the function ${\cal F}$ includes the
contribution coming from the exchange of the SM/LRSM $Z_1$ gauge
boson, ${\cal F}_1$ includes the contribution arising from the
exchange of the heavy gauge boson $Z_\theta$, while the function
$f$ contains the interference coming from both exchanges. Taking
the limit when $M_{Z_\theta}\to \infty$ and the mixing angle
$\phi=0$, the expressions for $C_V^{'e,\tau}$ and $C_A^{'e,\tau}$
reduce to $C_V^{'e,\tau}=C_A^{'e,\tau}=0$ and, Eq. (17) reduces to
the expression (4) given in Ref. \cite{Grifols} for the SM. On the
other hand, taking the limit when $M_{Z_\theta}\to \infty$ the Eq.
(17) reduces to the expressions (25) given in Ref.
\cite{A.Gutierrez} for the LRSM. Finally, if the mixing angle is
taken as $\phi=0$, Eq. (17) reduces to the expression for the
SM-$E_6$ models.

In the case of the weak dipole moments, to get the expression for
the differential cross section, we have to substitute  the $Z$ SM
couplings given in Eq. (8) by the respective weak dipole moments
included in Eq. (1), that is to say $a^W_\tau=F^Z_2(q^2=m^2_Z)$
and $d^W_\tau=eF^Z_3(q^2=m^2_Z)$. We do not reproduce the
analytical expressions here because they are rather similar to the
terms given in Eqs. (17-19). We applied a similar analysis
recently \cite{A.Gutierrez1} in order to get bounds on the MM and
EDM associated to the tau-neutrino using the L3 data obtained for
the reaction $e^{+}e^{-}\rightarrow \nu \bar\nu\gamma$ \cite{L3E}.
In the following section we will present the bounds obtained for
the tau dipole moments using the data published by the L3 and OPAL
Collaborations for the reaction $e^{+}e^{-}\rightarrow
\tau^+\tau^-\gamma$ \cite{L3,OPAL}.

\section{Results}

In practice, detector geometry imposes a cut on the photon polar
angle with respect to the electron direction, and further cuts
must be applied on the photon energy and minimum opening angle
between the photon and tau in order to suppress background from
tau decay products. In order to evaluate the integral of the total
cross section as a function of the parameters of the LRSM-$E_6$
models, that is to say, $\phi$, $M_{Z_\theta}$ and the mixing
angle $\theta_{E_6}$, we require cuts on the photon angle and
energy to avoid divergences when the integral is evaluated at the
important intervals of each experiment. We integrate over
$\cos\theta_\gamma$ from $-0.74$ to $0.74$ and $E_\gamma$ from 5
$GeV$ to 45.5 $GeV$ for various fixed values of the mixing angle
$\phi=-0.00166, 0, 0.00122$ (as is illustrate in Fig. 2) and for
$\theta_{E_6} = 37.8^o$ (which corresponds to $Z_\theta \to Z'$)
and $M_{Z_\theta}= 7M_{Z_1}$ according to Ref. \cite{Aytekin}.
Using the following numerical values: $\sin^2\theta_W=0.2314$,
$M_{Z_1}=91.18$ $GeV$, $\Gamma_{Z_\theta}=\Gamma_{Z_1}=2.49$ $GeV$
and
$z=(\frac{3}{5}\sin^2\theta_W)(\frac{M_{Z_1}}{M_{Z_\theta}})^2$ we
obtain the cross section $\sigma=\sigma(\phi, \theta_{E_6},
M_{Z_\theta}, a_\tau,d_\tau)$.

In Fig. 2, we show the dependence of the total cross section for
the process $e^+e^-\to \tau^+\tau^-\gamma$ with respect to the
LRSM mixing angle $\phi$. Using the limits obtained by the L3/OPAL
Collaborations \cite{L3,OPAL} for this process, we get the
following limits for $\phi$

\begin{equation}
-1.66\times 10^{-3}\leq \phi  \leq 1.22\times 10^{-3},
\end{equation}

\noindent which are consistent with those obtained recently from
the LEP data on the number of light neutrino species in the LRSM
\cite{A.Gutierrez2}, and they are about one order of magnitude
stronger than the previous limit obtained from old LEP data
\cite{M.Maya}. Since we have calculated the cross section at the
$Z_1$ pole, {\it i.e.} at $s=M^2_{Z_1}$, the value of
$\sin^2\theta_W$ is not affected by the $Z_\theta$ physics
\cite{Langacker1,Demir}. Variation of the $\Gamma_{Z_\theta}$ is
taken in the range from 0.15 to 2.0 times $\Gamma_{Z_1}$ in the
results of the CDF Collaboration \cite{CDF}. So we take
$\Gamma_{Z_\theta}=\Gamma_{Z_1}$ as a special case of this
variation.

As was discussed in Ref. \cite{L3}, $N\approx\sigma(\phi,
\theta_{E_6}, M_{Z_\theta}, a_\tau, d_\tau){\cal L}$, using
Poisson statistic \cite{L3,Barnett}, we require that
$N\approx\sigma(\phi, \theta_{E_6}, M_{Z_\theta}, \mu_{\nu_\tau},
d_{\nu_\tau}){\cal L}$ be less than 1559, with ${\cal L}= 100$
$pb^{-1}$, according to the data reported by the L3 Collaboration
Ref. \cite{L3} and references therein. Taking this into
consideration, we can get a bound for the tau magnetic moment as a
function of $\phi$, $\theta_{E_6}$ and $M_{Z_\theta}$ with
$d_\tau=0$. The values obtained for this bound for several values
of $\phi$ with $\theta_{E_6}=37.8^o$ and $M_{Z_\theta}=7M_{Z_1}$
are included in Table 1. The previous analysis and comments can
readily be translated to the EDM of the tau with $a_\tau=0$. The
resulting bounds for the EDM as a function of $\phi$,
$\theta_{E_6}$ and $M_{Z_\theta}$ are shown in Table 1. As
expected, the limits obtained for the electromagnetic dipole
moments of the tau lepton are consistent with those obtained by
the these collaborations from the data obtained for the process
$e^+e^-\to \tau^+\tau^-\gamma$ \cite{L3,OPAL}.

\begin{center}
\begin{tabular}{|c|c|c|}\hline
$\phi$&${a_\tau}$&$d_\tau(10^{-16}e \mbox{cm})$\\
\hline \hline
$-1.66\times 10^{-3}$&0.0521&2.891\\
\hline
0&0.052&2.885\\
\hline
$1.22\times 10^{-3}$&0.0519&2.881\\
\hline
\end{tabular}
\end{center}

\begin{center}
Table 1. Limits on the $a_\tau$ MM and $d_\tau$ EDM of the
$\tau$-lepton for different values of the mixing angle $\phi$ with
$\theta_{E_6}=37.8^o$ and $M_{Z_\theta}=7M_{Z_1}$. We have applied
the cuts used by L3 for the photon angle and energy.
\end{center}

\vspace{3mm}

The bounds for the weak dipole moments of the tau-lepton according
to the data from the L3 and OPAL Collaboration \cite{L3,OPAL} for
the energy and the opening angle of the photon, as well as the
luminosity and the events numbers are given in the Table 2. As we
can appreciate, the use of the strong limit obtained for the
$\phi$ mixing angle induces also stringet bounds for the tau weak
dipole moments, which are already consistent with those bound
recently by the DELPHI and ALEPH Collaborations in the process $
e^+e^-\to \tau^+\tau^-$ \cite{DELPHI,ALEPH}.

\begin{center}
\begin{tabular}{|c|c|c|}\hline
$\phi$&${a^W_\tau(10^{-3}) }$&$d^W_\tau(10^{-17}e \mbox{cm})$\\
\hline \hline
$-1.66\times 10^{-3}$&2.143&1.19\\
\hline
0&2.138&1.187\\
\hline
$1.22\times 10^{-3}$&2.135&1.185\\
\hline
\end{tabular}
\end{center}

\begin{center}
Table 2. Limits on the $a^W_\tau$ anomalous weak MM and $d^W_\tau$
weak EDM of the $\tau$-lepton for different values of the mixing
angle $\phi$. We have applied the cuts used by L3 for the photon
angle and energy.
\end{center}

As far as the respective analysis for the $E_6$ model is
concerned, we will use the known limits for the mixing angle
$\phi_{(SM-E_6)}$ \cite{London,Capstick},

\begin{equation}
-0.054\leq \phi_{(SM-E_6)} \leq 0.054,
\end{equation}

\noindent in order to get the respective limits of the tau dipole
moments. In Tables 3 and 4 we show these limits. Here we obtain
also limits for the electromagnetic moments that are consistent
with the old results published by the L3/OPAL Collaborations
\cite{L3,OPAL}, but our limits for the weal dipole moments are
consistent with the data published recently by the DELPHI/ALEPH
Collaborations.

\begin{center}
\begin{tabular}{|c|c|c|}\hline
$\phi_{(SM-E_6)}$&${a_\tau}$&$d_\tau(10^{-16}e \mbox{cm})$\\
\hline \hline
-0.054&0.047&2.64\\
\hline
0&0.052&2.88\\
\hline
0.054&0.054&2.92\\
\hline
\end{tabular}
\end{center}

\begin{center}
Table 3. Limits on the $a_\tau$ MM and $d_\tau$ EDM of the
$\tau$-lepton for different values of the mixing angle
$\phi_{(SM-E_6)}$ with $\theta_{E_6}=37.8^o$ and
$M_{Z_\theta}=7M_{Z_1}$. We have applied the cuts used by L3 for
the photon angle and energy.
\end{center}

\begin{center}
\begin{tabular}{|c|c|c|}\hline
$\phi_{(SM-E_6)}$&${a^W_\tau(10^{-3}) }$&$d^W_\tau(10^{-17}e \mbox{cm})$\\
\hline \hline
-0.054&1.91&1.08\\
\hline
0&2.13&1.18\\
\hline
0.054&2.20&1.22\\
\hline
\end{tabular}
\end{center}

\begin{center}
Table 4. Limits on the $a^W_\tau$ anomalous weak MM and $d^W_\tau$
weak EDM of the $\tau$-lepton for different values of the mixing
angle $\phi_{(SM-E_6)}$. We have applied the cuts used by L3 for
the photon angle and energy.
\end{center}

We plot the total cross section in Fig. 3 as a function of the
mixing angle $\phi$ for the bounds of the magnetic moment given in
Table 1 with $\theta_{E_6}=37.8^0$ and $M_{Z_\theta}=7M_{Z_1}$. We
reproduce the Fig. 2 of the Ref. \cite{A.Gutierrez}. Our results
for the dependence of the differential cross section on the photon
energy versus the cosine of the opening angle between the photon
and the beam direction ($\theta_\gamma$) are presented  in Fig. 4,
for $\phi = 0.00122$ and $a_\tau = 0.0519$. In this case also we
reproduce the Fig. 3 of the Ref. \cite{A.Gutierrez}. Finally we
plot the differential cross section in Fig. 5 as a function of the
photon energy for the bounds of the magnetic moments given in
Table 1. We observed in this figure that the energy distributions
are consistent with those reported in the literature.

In a previous paper \cite{A.Gutierrez} we estimated bounds on the
anomalous magnetic moment and the electric dipole moment of the
tau through the process $e^{+}e^{-}\rightarrow \tau^+\tau^-
\gamma$ in the context of a LRSM at the $Z_1$ pole. We found that
these bound were almost independent of the mixing angle $\phi$ of
the model. In the present paper we reproduce these bounds for
$\theta_{E_6}=37.8^o$ and $M_{Z_\theta}=7M_{Z_1}$ \cite{Aytekin},
corresponding to the $E_6$ superstring models. Our results in
Table 2 for $\phi=-0.00166, 0, 0.00122$ and Table 4 for
$\phi_{(SM-E_6)}=-0.054, 0, 0.054$ with $\theta_{E_6}=37.8^o$ and
$M_{Z_\theta}=7M_{Z_1}$ confirm the bounds obtained by the DELPHI
and ALEPH collaborations for the weak tau dipole moments
\cite{Lohmann,DELPHI,ALEPH}; however, our analysis is not
sensitive to the real and imaginary parts of these parameters
separately. On the other hand, it seems that in order to improve
these limits it might be necessary to study direct CP-violating
effects \cite{M.A.Perez,Larios}.

\section{Conclusions}

We have determined limits on the electromagnetic and weak dipole
moments of the tau lepton using the data published by the L3 and
OPAL Collaborations for the process $e^+e^-\to
\tau^+\tau^-\gamma$. We were able to get strong limits on the weak
dipole moments by constraining the LRSM mixing angle $\phi$ from
the electromagnetic dipole moments obtained by these
collaborations. We then used this limit to constrain the weak
dipole moments of the tau lepton from the same L3/OPAL data and we
have obtained similar bounds for these dipole moments as those
obtained recently by the DELPHI/ALEPH  Collaborations from the
process $e^+e^-\to \tau^+\tau^-$ \cite{DELPHI,ALEPH}. We have
obtained similar bounds in the case of the $E_6$ string models for
appropriated values of its parameters. In particular, in the limit
$\phi=0$, $\phi_{(SM-E_6)}=0$ and $M_{Z_\theta}\to \infty$, our
bound takes the value previously reported in Ref. \cite{Grifols}
for the SM. The bounds in the MM and the EDM are not affected for
the additional neutral vector boson $Z_\theta$ since its mass is
higher than $Z_1$ at $\sqrt{s}=M_{Z_1}$. But at higher
center-of-mass energies $\sqrt{s}\sim M_{Z_\theta}$, the
$Z_\theta$ contribution to the cross section becomes comparable
with $Z_1$. As far as the weak dipole moments are concerned, our
limits given in Tables 1-4 are consistent with the experimental
bounds obtained at LEP with the two-body decay mode $Z_1 \to
\tau^+ \tau^-$ \cite{Lohmann}. In addition, the analytical and
numerical results for the total cross section have never been
reported in the literature before and could be of relevance for
the scientific community.

\vspace{1cm}

\begin{center}
{\bf Acknowledgments}
\end{center}

We acknowledge support from CONACyT and SNI (M\'exico).

\newpage

\begin{figure}[t]
\centerline{\scalebox{0.85}{\includegraphics{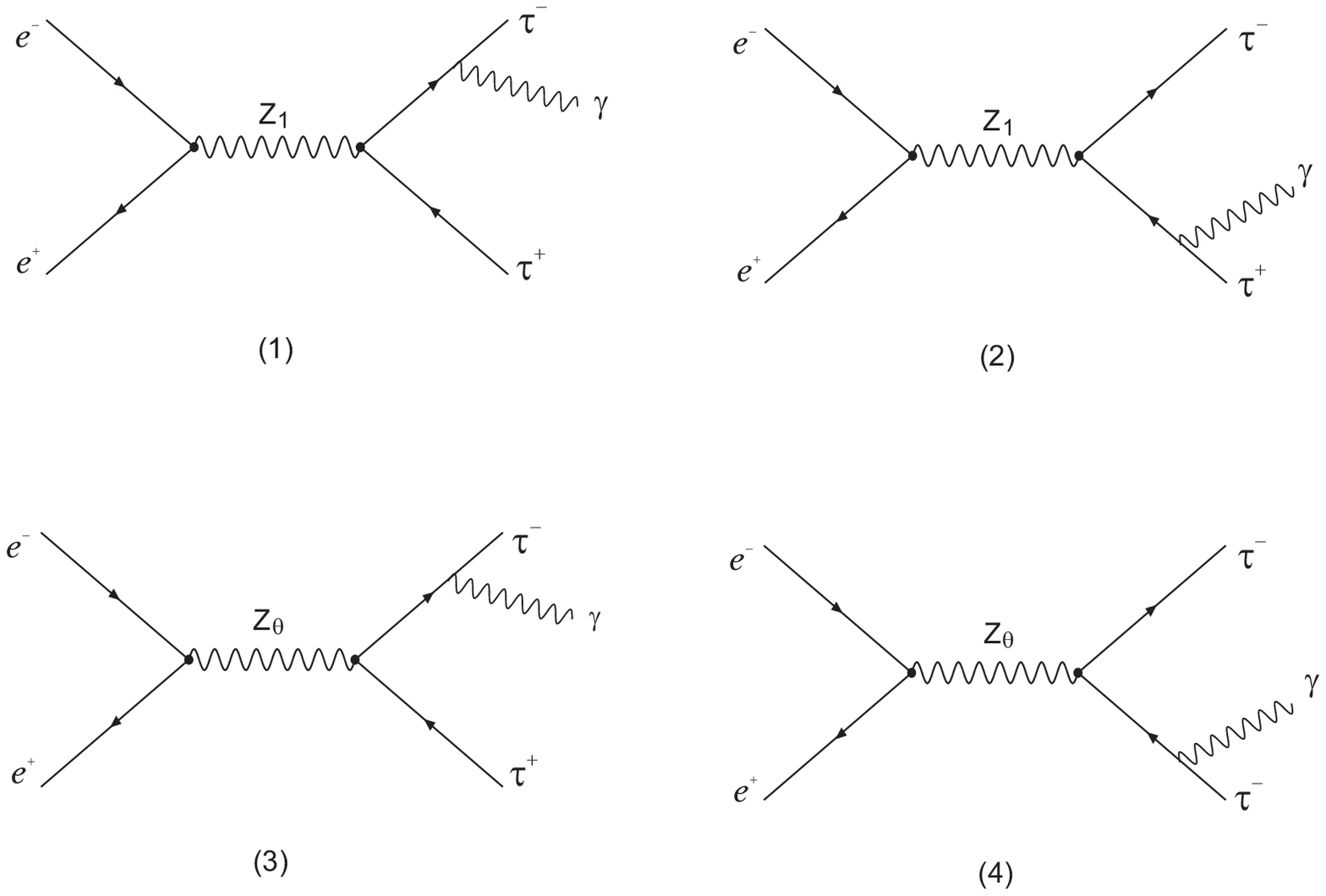}}}
\caption{ \label{fig:gamma} The Feynman diagrams contributing to
the process $e^{+}e^{-}\rightarrow \tau^+\tau^-\gamma$ in a
left-right symmetric model, and in the $E_6$ model.}
\end{figure}

\begin{figure}[t]
\centerline{\scalebox{0.75}{\includegraphics{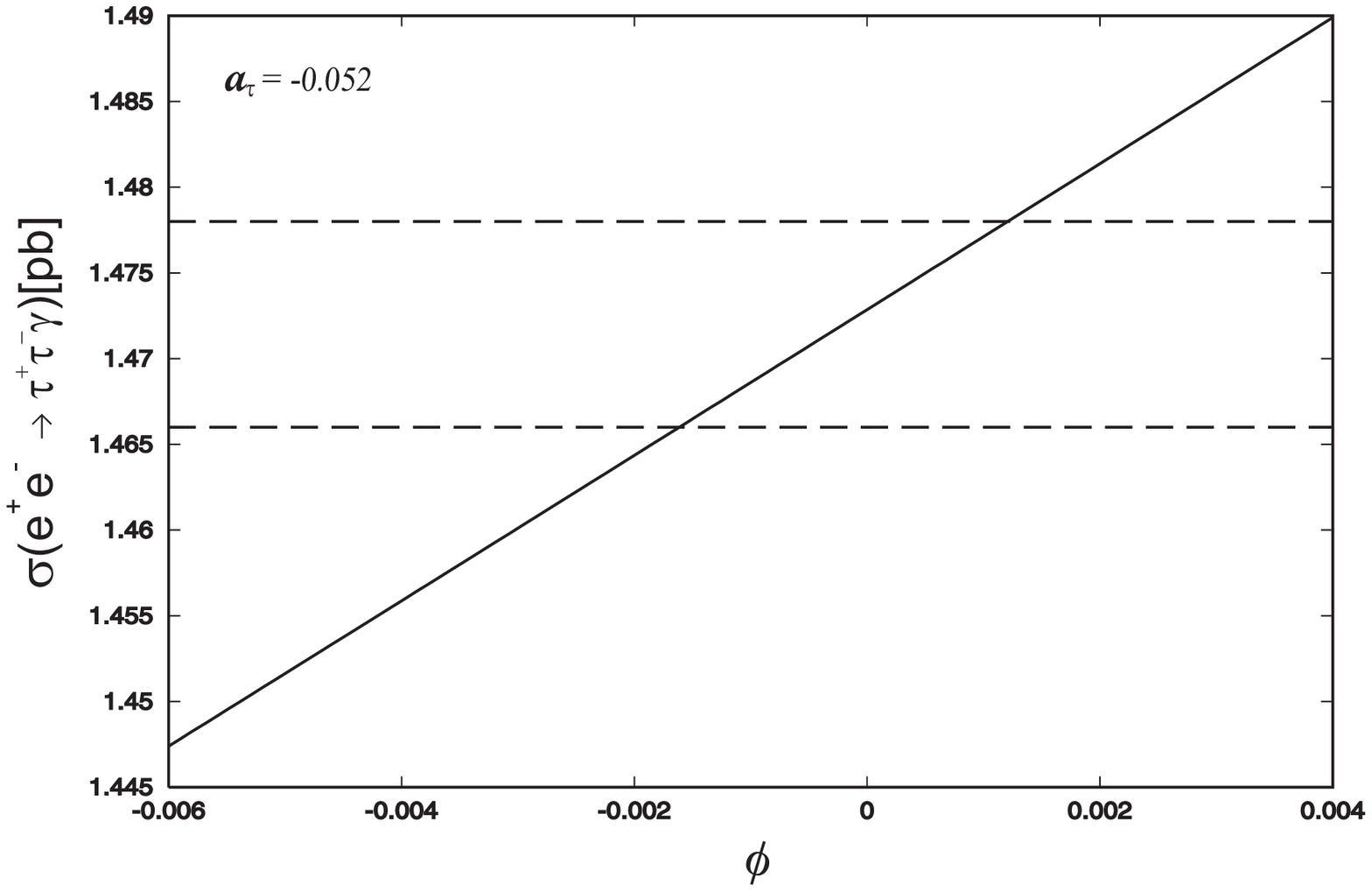}}}
\caption{ \label{fig:gamma} The curves show the shape for
$\sigma(e^{+}e^{-}\rightarrow \tau^+\tau^-\gamma)$ as a function
of the mixing angle $\phi$, with $a_\tau =-0.052$.}
\end{figure}

\begin{figure}[t]
\centerline{\scalebox{0.85}{\includegraphics{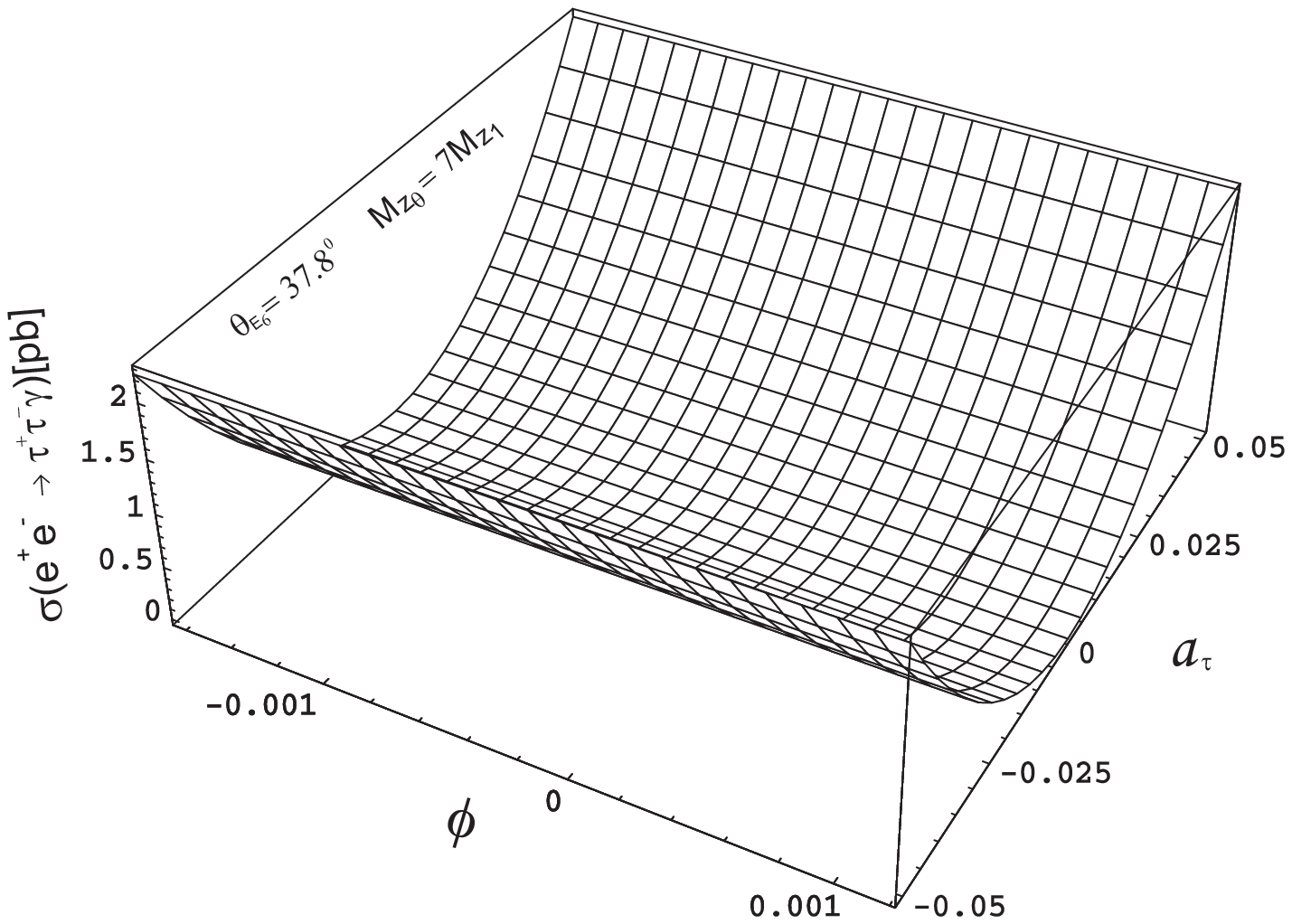}}}
\caption{ \label{fig:gamma} The total cross section for
$e^{+}e^{-}\rightarrow \tau^+\tau^-\gamma$ as a function of $\phi$
and $a_\tau$ (Table 1) with $\theta_{E_6}=37.8^o$ and
$M_{Z_\theta}=7M_{Z_1}$.}
\end{figure}

\begin{figure}[t]
\centerline{\scalebox{0.85}{\includegraphics{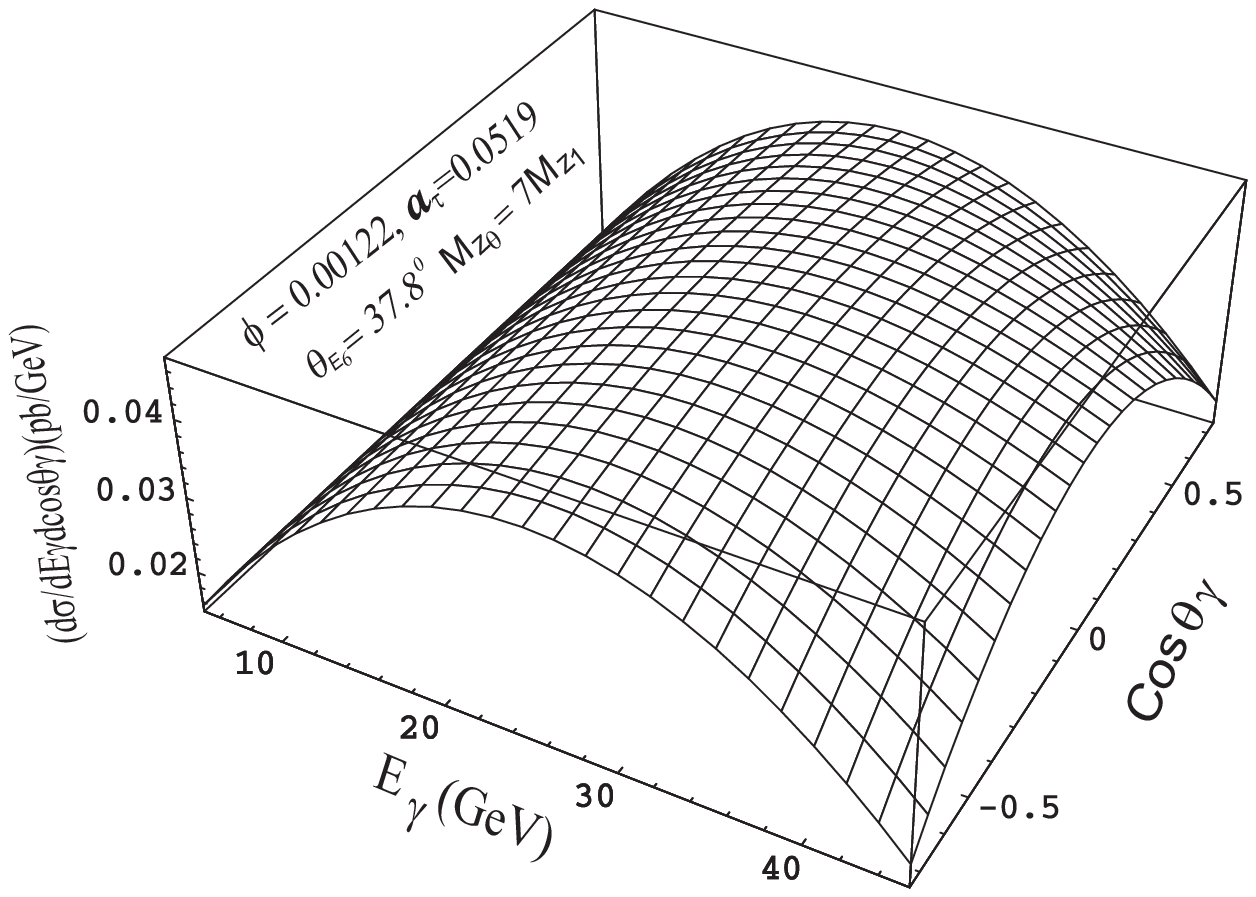}}}
\caption{ \label{fig:gamma} The differential cross section for
$e^{+}e^{-}\rightarrow \tau^+\tau^-\gamma$ as a function of
$E_\gamma$ and $\cos\theta_\gamma$ for $\phi=1.22\times 10^{-3}$,
$a_\tau=0.0519$, $\theta_{E_6}=37.8^o$ and $M_{Z_\theta}=7M_{Z_1}$
.}
\end{figure}

\begin{figure}[t]
\centerline{\scalebox{0.85}{\includegraphics{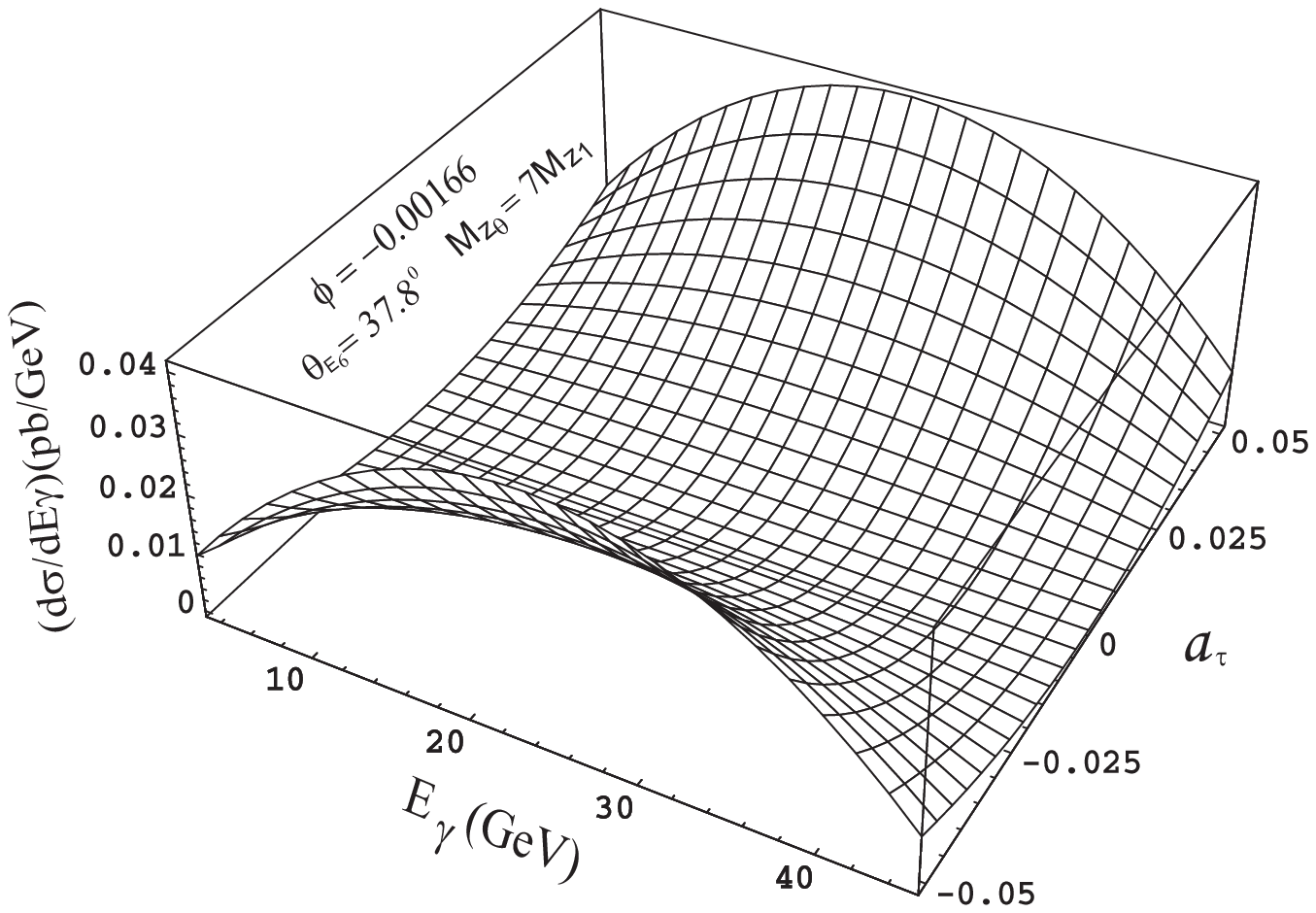}}}
\caption{ \label{fig:gamma} The differential cross section for
$e^{+}e^{-}\rightarrow \tau^+\tau^-\gamma$ as a function of
$E_\gamma$ and $a_\tau$ with $\phi=-1.66\times 10^{-3}$,
$\theta_{E_6}=37.8^o$ and $M_{Z_\theta}=7M_{Z_1}$.}
\end{figure}

\end{document}